\documentstyle[a4,12pt]{article}

\title{Equations of the reaction--diffusion type with a loop algebra 
       structure}

\author{E. Alfinito, V. Grassi, R. A. Leo, G. Profilo and G. Soliani \\
     Dipartimento di Fisica dell'Universit\`a,\\
     73100 Lecce, Italy,\\
     and Istituto Nazionale di Fisica Nucleare,\\
     Sezione di Lecce, Italy}

\begin{document}

\pagebreak
\maketitle
\begin{abstract}

A system of equations of the reaction--diffusion type is studied in the 
framework of both the direct and the inverse prolongation structure. We 
find that this system allows an incomplete prolongation Lie algebra, which 
is used to find the spectral problem and a whole class of nonlinear field 
equations containing the original ones as a special case.

\end{abstract}

\section{Introduction}

Mathematical models of spatial pattern formation in complex organisms are 
based on nonlinear interactions of at least two chemicals and on their 
diffusion, where autocatalysis and long--range inhibition play a crucial 
role \cite{K}. Generally, these models are described by nonlinear 
evolution equations of the reaction--diffusion type, a class of which can 
be written as
\begin{eqnarray}
& u_t=D_1\Delta u+b_1 u^2 v+b_2 uv^2+b_3 u+b_4 v+b_5,\label{A1a}\\
& v_t=D_2\Delta v+c_1 u^2 v+c_2 uv^2+c_3 u+c_4 v+c_5,\label{A1b}
\end{eqnarray}
where $u=u(x,\,y,\,t)$, $v=v(x,\,y,\,t)$, $\Delta$ denotes the Laplace 
operator in two--dimensional orthonormal coordinates, $D_1$, $D_2$ are the 
diffusion constants, and $b_1$, $c_1,\ldots$ are (constant) coefficients.
\par
The physical meaning of $u$, $v$ and the parameters $b_1$, $c_1,\ldots$ 
depends on the biological phenomenon under consideration. For example, in 
the case of the activator--substrate model in its simplest form \cite{K}, 
one has $b_1=-b_3$, $b_2=b_4=b_5=0$, $c_5=-c_1$, $c_2=c_3=c_4=0$, where 
$b_1$ and $c_1$ mean the cross-reaction coefficients. The functions $u$ and 
$v$ can be interpreted as the self--enhanced reactant and a substrate 
depleted by $u$, respectively.\par
The high interest presented by the mathematical formulation of pattern 
generation in complex structures rises the question whether the system 
(\ref{A1a})--(\ref{A1b}) contains integrable subcases. Due to the 
hardness of the subject, it is convenient to dwell upon this problem 
gradually, i.e. first studying a (1+1)--dimensional version of Eqs.
(\ref{A1a})--(\ref{A1b}). To this regard, we notice that biological systems 
modeled by equations belonging to the class (\ref{A1a})--(\ref{A1b}) are 
interesting also in (1+1)--dimensions. One of them, which consists of Eqs. 
(\ref{A1a})--(\ref{A1b}) where the nonlinear terms are missing, was 
introduced by Kondo and Asai in 1995 in relation to the study of the stripe 
pattern of the angelfish {\it{Pomacanthus}} {\cite{KA}}.\par
In performing our programme, we remind the reader that one of the most 
remarkable feature of integrable nonlinear field equations is the onset of 
infinite dimensional Lie algebras with a loop structure. This characteristic 
strongly suggests that a close connection should be estabilished between the 
integrability property and the infinite dimension of the Lie algebra allowed 
by a given nonlinear field equation. An efficacious tool to look for the 
existence of loop algebras for 1+1 dimensional nonlinear systems is based on 
the Estrabrook--Wahlquist (EW) prolongation theory \cite{WE}. The application 
of this procedure to Eqs. (\ref{A1a})--(\ref{A1b}) in 1+1 dimensions tells us 
that the system
\begin{eqnarray}
& u_t-u_{xx}+2u^2v-2ku=0,\label{A2a}\\
& v_t+v_{xx}-2uv^2+2kv=0,\label{A2b}
\end{eqnarray}
where $k$ is a constant, is endowed with a Lie algebra possessing a loop 
structure.\par
It is noteworthy that Eqs. (\ref{A2a})--(\ref{A2b}) emerge also in the 
gauge formulation of the 1+1 dimensional gravity \cite{MPS}. In this context, 
$u$ and $v$ are Zweibein fields. In general, the system 
(\ref{A2a})--(\ref{A2b}) is similar to the "fictitious" or "mirror--image" 
systems with {\it negative} friction, which appear into the thermo--field 
approach to the damped oscillator treated in \cite{CRV}.\par
Other equations which fall into the class (\ref{A1a})--(\ref{A1b}) admitting 
nontrivial Lie algebras, although of finite dimensions, will be handled 
elsewhere.\par
Here we study systematically Eqs. (\ref{A2a})--(\ref{A2b}) in both the 
direct and the inverse prolongation framework. The inverse prolongation 
method consists in starting from a given incomplete Lie algebra to find the 
nonlinear field equations whose prolongation structure it is.\par
The direct prolongation method is carried out is Sec. 2. It provides an 
incomplete Lie algebra (in the sense that not all of the commutators are 
known) which is exploited to obtain the linear eigenvalue problem 
associated with the system (\ref{A2a})--(\ref{A2b}). A possible 
realization of the prolongation algebra turns out to be an 
infinite--dimensional Lie algebra with a loop structure of the Kac--Moody 
type. In Sec. 3 we deal with the inverse prolongation. We consider the 
incomplete Lie algebra determined in Sec. 2 to build up the differential 
ideal related to Eqs. (\ref{A2a})--(\ref{A2b}). In Sec. 4 an inverse 
prolongation procedure based on a certain realization of the Kac--Moody 
algebra related to Eqs. (\ref{A2a})--(\ref{A2b}) is outlined. In both these 
inverse schemes we can generate the class of field equations whose 
prolongation structure the incomplete Lie algebra is. New integrable equations 
arise including the original ones as a special case. Finally, in Sec. 5 some 
comments are reported while the Appendixes A, B and C contain details of the 
calculations.

\section{The prolongation algebra}

In order to formulate the EW prolongation method for Eqs. 
(\ref{A2a})--(\ref{A2b}) let us introduce the differential ideal defined by 
the set of 2--forms
\begin{eqnarray}
&\alpha_1=du\wedge dt-u_x dx\wedge dt,\label{B1a}\\
&\alpha_2=dv\wedge dt-v_x dx\wedge dt,\label{B1b}\\
&\alpha_3=-du\wedge dx-du_x\wedge dt+2u(uv-k)dx\wedge dt,\label{B1c}\\
&\alpha_4=-dv\wedge dx+dv_x\wedge dt-2v(uv-k)dx\wedge dt,\label{B1d}
\end{eqnarray}
where $\wedge$ means the wedge product. The ideal 
(\ref{B1a})--(\ref{B1d}) turns out to be closed.\par
Now we consider the prolongation 1--forms
\begin{equation}
\omega^k=dy^k+F^k(u,\,u_x,\,v,\,v_x;\,y)dx+G^k(u,\,u_x,\,v,\,v_x;\,y)dt,
\label{B2}
\end{equation}
where $y=\{ y^m\}$, $k$, $m=1,\,2,\,\ldots,N$ ($N$ arbitrary), and $F^k$, 
$G^k$ are, respectively, the pseudopotential and functions to be 
determined. By requiring that 
$d\omega^k\in{\cal I}(\alpha_j,\,\omega^k)$,
${\cal I}$ being the ideal generated by $\alpha_j$ and $\omega^k$, we find 
the constraints
\begin{eqnarray}
& F_u^k-G_{u_x}^k=0,\label{B3a}\\
& F_{u_x}^k=0,\label{B3b}\\
& F_v^k+G_{v_x}^k=0,\label{B3c}\\
& F_{v_x}^k=0,\label{B3d}\\
& 2u(uv-k)F_u^k-2v(uv-k)F_v^k+G_u^ku_x+G_v^kv_x+[F,\,G]^k=0,\label{B3e}
\end{eqnarray}
where $[F,\,G]^k=F^jG_{y_j}^k-G^jF_{y_j}^k$, 
$F_{y_j}^k={{\partial F^k}\over{\partial y_j}}$,
$F_u^k={{\partial F^k}\over{\partial u}}$, and so on.\par
In the following, we shall omit the index $k$, for simplicity.\par
The differential equations (\ref{B3a})--(\ref{B3e}) yield
\begin{eqnarray}
& F=a_1 uv+a_2 u+a_3 v+a_4,\label{B4a}\\
& G=(a_1 v+a_2)u_x-(a_1 u+a_3)v_x+[a_3,\,a_2]uv+\nonumber\\
&[a_4,\,a_2]u+[a_3,\,a_4]v+a_5,\label{B4b}
\end{eqnarray}
and the commutation relations
\begin{eqnarray}
& [a_1,a_2]=0,\;\;\;[a_1,a_3]=0,\;\;\;[a_1,a_4]=0,\;\;\;
[a_4,a_5]=0,\label{B5a}\\
& [a_1,[a_3,a_2]]=0,\;\;\;[a_1,[a_3,a_4]]=0,\;\;\;
[a_1,[a_4,a_2]]=0,\label{B5b}\\
& [a_2,[a_3,a_2]]=2a_2,\;\;\;[a_2,[a_4,a_2]]=0,\;\;\;
[a_3,[a_3,a_2]]=-2a_3\label{B5c}\\
& [a_3,[a_3,a_4]]=0,\;\;\;2[a_4,[a_3,a_2]]+[a_1,a_5]=0,\label{B5d}\\
& 2ka_2+[a_2,a_5]+[a_4,[a_4,a_2]]=0,\label{B5e}\\
& -2ka_3+[a_3,a_5]+[a_4,[a_3,a_4]]=0.\label{B5f}
\end{eqnarray}
Here $a_j$ $(j=1,\,2,\,\ldots,5)$ are functions of integration depending on 
the pseudopotential $y$ only.\par
Equations (\ref{B5a})--(\ref{B5f}) define an incomplete Lie algebra. 
However, we can find a homomorphism between the algebra 
(\ref{B5a})--(\ref{B5f}) and the $sl(2,{\cal R})$ algebra
\begin{equation}
[X_1,X_2]=2X,\;\;\;[X_2,X]=2X_1,\;\;\;[X,X_1]=-2X_2,
\label{B6}
\end{equation}
where $X=[a_3,a_2]$, $X_1=a_2+a_3$ and $X_2=a_2-a_3$. This can be seen 
assuming that
\begin{equation}
a_1=0,\;\;\;a_4=\lambda X,\;\;\;a_5=-(k+2\lambda^2)X,
\label{B7}
\end{equation}
where $\lambda$ is a free parameter. Then, Eqs. (\ref{B5a})--(\ref{B5f}) 
yield easily (\ref{B6}).\par
By means of (\ref{B6}), we can build up the spectral problem related to 
Eqs. (\ref{A2a})--(\ref{A2b}). In doing so, we need to exploit a 
realization of Eqs. (\ref{B6}). In terms of two prolongation variables 
$(y_1,y_2)$ a linear realization of (\ref{B6}) is given by
\begin{equation}
X_1=y_2\partial_{y_1}+y_1\partial_{y_2},\;\;\;
X_2=y_2\partial_{y_1}-y_1\partial_{y_2},\;\;\;
X=y_1\partial_{y_1}-y_2\partial_{y_2},
\label{B8}
\end{equation}
to which the following $2\times 2$ matrix representation
\begin{equation}
X_1=\left(\begin{array}{cc}
                 0 & 1 \\
                 1 & 0 
          \end{array} 
                     \right),\;\;\;
X_2=\left( 
          \begin{array}{cc}
                 0 & -1 \\
                 1 &  0 
          \end{array} 
                     \right),\;\;\;
X=\left( 
        \begin{array}{cc}
               1 &  0 \\
               0 & -1 
        \end{array} 
                   \right),
\label{B9}
\end{equation}
corresponds. Keeping in mind (\ref{B2}), (\ref{B4a}) and (\ref{B4b}), with 
the help of (\ref{B9}) we obtain the spectral problem for Eqs. 
(\ref{A2a})--(\ref{A2b}), namely
\begin{equation}
{\bf y}_x=-F^T{\bf y},\;\;\;{\bf y}_t=-G^T{\bf y},
\label{B10}
\end{equation}
where ${\bf y}=(y_1,\,y_2)^T$, 
\begin{equation}
F=\left(
        \begin{array}{cc}
               \lambda &  v       \\
                  u    & -\lambda 
        \end{array} 
                   \right),\;\;\;
G=\left(
        \begin{array}{cc}
               uv-k-2\lambda^2 & -(v_x+2\lambda v) \\
               u_x-2\lambda u  & -uv+k+2\lambda^2
        \end{array} 
                   \right).
\label{B11}
\end{equation}
The compatibility condition for Eqs. (\ref{B10}), i.e. 
$L_{1t}-L_{2x}-[L_1,L_2]=0$, provides just Eqs. (\ref{A2a})--(\ref{A2b}).\par
Another possible realization of the prolongation algebra 
(\ref{B5a})--(\ref{B5f}) is given by an infinite--dimensional Lie algebra 
with a loop structure, precisely an algebra of the Kac--Moody type. To this 
aim, let us suppose that $a_1=0$. Then let us put
\begin{equation}
a_2=\stackrel{(m_0)}{T_1}-i\stackrel{(-m_0)}{T_2},
\label{B12a}
\end{equation}
\begin{equation}
a_3=\stackrel{(m_0)}{T_1}+i\stackrel{(-m_0)}{T_2},
\label{B12b}
\end{equation}
\begin{equation}
a_4=\alpha_4\stackrel{(n_4)}{T_3}+\beta_4\stackrel{(-n_4)}{T_3},
\label{B12c}
\end{equation}
\begin{equation}
a_5=\alpha_5\stackrel{(n_5)}{T_3}+\beta_5\stackrel{(-n_5)}{T_3},
\label{B12d}
\end{equation}
where $m_0$, $n_4$, $n_5 \in {\cal Z}$, and the coefficients 
$\alpha_j,\;\beta_j\in {\cal{R}}\;\;(j=4,\,5)$, are to be determined in such a 
way that (\ref{B12a})--(\ref{B12d}) satisfy the commutation relations of the 
(incomplete) Lie algebra (\ref{B5a})--(\ref{B5f}). A straightforward 
calculation shows that the vector fields 
$\stackrel{(n)}{T_j}\;(j=1,\,2,\,3;\;n\in{\cal Z})$
are elements of the Kac--Moody algebra (without central charge) governed by 
the commutation relations
\begin{equation}
[\stackrel{(n)}{T_j},\stackrel{(m)}{T_k}]=i\epsilon_{jkl}
\stackrel{(n+m)}{T_l}
\label{B13}
\end{equation}
$(n,\,m\in{\cal Z})$, $\epsilon_{jkl}$ being the Ricci tensor. The 
requirement that Eqs. (\ref{B5a})--(\ref{B5f}) are fulfilled, entails 
$m_0=0,$ $n_5=2n_4$, $\alpha_4=-\beta_4=\sqrt{k}$ $(k>0)$, and $\alpha_5=
\beta_5=-k$.\par
The algebra (\ref{B13}) admits the following realization in terms of 
the prolongation variables:
\begin{equation}
\stackrel{(m)}{T_1}={1\over 2}\sum_{n\in {\cal Z}}
\left[ y_2^{(m+n)}{\partial\over {\partial y_1^{(n)}}}+
y_1^{(m+n)}{\partial\over {\partial y_2^{(n)}}}\right],
\label{B14a}
\end{equation}
\begin{equation}
\stackrel{(m)}{T_2}={i\over 2}\sum_{n\in {\cal Z}}
\left[ y_2^{(m+n)}{\partial\over {\partial y_1^{(n)}}}-
y_1^{(m+n)}{\partial\over {\partial y_2^{(n)}}}\right],
\label{B14b}
\end{equation}
\begin{equation}
\stackrel{(m)}{T_3}={1\over 2}\sum_{n\in {\cal Z}}
\left[ y_1^{(m+n)}{\partial\over {\partial y_1^{(n)}}}-
y_2^{(m+n)}{\partial\over {\partial y_2^{(n)}}}\right],
\label{B14c}
\end{equation}
where the pseudopotential $y$ is expressed via the independent 
infinite--dimensional vectors $y_1$ and $y_2$ ($y_1^{(n)}$ and $y_2^{(n)}$ 
stand for the $n$--component of $y_1$ and $y_2$, respectively).\par
At this point it is instructive to exploit the infinite--dimensional 
algebra (\ref{B13}) to write down again the spectral problem related to 
Eqs. (\ref{A2a})--(\ref{A2b}). The procedure, which is outlined in Appendix 
A, is based on the equations $y_x=-F$, $y_t=-G$ (see (\ref{B2})), where $F$ 
and $G$ are given by (\ref{B4a})--(\ref{B4b}) with $a_1=0$, while $a_2$, 
$a_3$, $a_4$ and $a_5$ are expressed in terms of the Kac--Moody operators by 
(\ref{B12a})--(\ref{B12d}), using the representation 
(\ref{B14a})--(\ref{B14c}).

\section{The inverse prolongation}

In Sec. 2, we have applied the E--W method to find the prolongation algebra 
associated with Eqs. (\ref{A2a})--(\ref{A2b}). In general, the prolongation 
of an integrable nonlinear field equation can be interpreted as a 
Cartan--Ehresmann connection, so that an incomplete Lie algebra of vectors 
field can be related to a differential ideal \cite{E}. On the contrary, 
starting from the incomplete Lie algebra (\ref{B5a})--(\ref{B5f}), we can 
yield the differential ideal associated with Eqs. (\ref{A2a})--(\ref{A2b}) 
specifying the form of the connection. In this way, we can generate the field 
equations whose prolongation structure the incomplete algebra 
(\ref{B5a})--(\ref{B5f}) is. To this aim, let us assume that the connection
\begin{equation}
\omega^k=dy^k+A_j^k\theta^j
\label{C1}
\end{equation}
exists, such that
\begin{equation}
d\omega^k=A_j^kd\theta^j-{1\over 2}[A_j,\,A_i]^k\theta^j\wedge
\theta^i\;\;\;({\rm mod}\;\omega^k),
\label{C2}
\end{equation}
where $\theta^j$ are 1--forms, $A_j$ $(j=1,\,2,\ldots,8)$ are defined by
\begin{eqnarray}
&A_1=a_1,\;\;\;A_2=a_2,\;\;\;A_3=a_3,\;\;\;A_4=a_4,\;\;\;A_5=a_5,
\nonumber\\
&A_6=[a_3,\,a_2],\;\;\;A_7=[a_2,\,a_4],\;\;\;A_8=[a_3,\,a_4],
\label{C3}
\end{eqnarray}
appearing in the incomplete Lie algebra (\ref{B5a})--(\ref{B5f}), and 
${\rm{mod}}\,\omega^k$ means that all the exterior products between $\omega^k$ 
and 1--forms of the Grassmann algebra have not been considered. (For the 
reader convenience, in Appendix B we report the incomplete algebra involved 
in ({\ref{C2}) where one can easily recognize the independent commutation 
relations).\par
By resorting to the commutation relations of Appendix B, Eq. (\ref{C2}) 
provides the constraints
\begin{eqnarray}
&d\theta^1=0,\label{C4}\\
&d\theta^2+2k\theta^2\wedge\theta^5-2\theta^2\wedge\theta^6=0,\label{C5}\\
&d\theta^3-2k\theta^3\wedge\theta^5+2\theta^3\wedge\theta^6=0,\label{C6}\\
&d\theta^4=0,\label{C7}\\
&d\theta^5=0,\label{C8}\\
&d\theta^6+\theta^2\wedge\theta^3=0,\label{C9}\\
&d\theta^7-\theta^2\wedge\theta^4+2\theta^6\wedge\theta^7=0,\label{C10}\\
&d\theta^8-\theta^3\wedge\theta^4-2\theta^6\wedge\theta^8=0,\label{C11}\\
&2\theta^1\wedge\theta^5+\theta^3\wedge\theta^7-
  \theta^4\wedge\theta^6=0,\label{C12}\\
&\theta^2\wedge\theta^5+\theta^4\wedge\theta^7=0,\label{C13}\\
&\theta^2\wedge\theta^8+\theta^3\wedge\theta^7=0,\label{C14}\\
&\theta^3\wedge\theta^5-\theta^4\wedge\theta^8=0,\label{C15}\\
&\theta^5\wedge\theta^6=\theta^5\wedge\theta^7=
  \theta^5\wedge\theta^8=\theta^7\wedge\theta^8=0\label{C16}.
\end{eqnarray}\par
In the following, we shall consider an interesting solution of Eqs. 
(\ref{C4})--(\ref{C16}). To this aim, we note that Eq. (\ref{C8}) is 
satisfied if
\begin{equation}
\theta^5=dt.
\label{C17}
\end{equation}
On the other hand, Eq. (\ref{C16}) yields
\begin{equation}
\theta^6=\gamma_1 dt,\;\;\;\theta^7=\gamma_2 dt,\;\;\;
\theta^8=\gamma_3 dt,
\label{C18}
\end{equation}
where $\gamma_j$ $(j=1,\,2,\,3)$ are (arbitrary) 0--forms. Furthermore, 
Eqs. (\ref{C13})--(\ref{C15}) imply
\begin{eqnarray}
&\theta^2+\gamma_2\theta^4=\eta_1 dt,\label{C19}\\
&\gamma_3\theta^2+\gamma_2\theta^3=\eta_2 dt,\label{C20}\\
&\theta^3-\gamma_3\theta^4=\eta_3 dt,\label{C21}
\end{eqnarray}
where $\eta_j$ $(j=1,\,2,\,3)$ are 0--forms.\par
With the help of (\ref{C17}) and (\ref{C18}), Eq. (\ref{C12}) gives
\begin{equation}
2\theta^1+\gamma_2\theta^3-\gamma_1\theta^4=\eta_4 dt,
\label{C22}
\end{equation}
$\eta_4$ being a 0--form.\par
Now, from Eqs. (\ref{C19})--(\ref{C21}) we obtain
\begin{equation}
\gamma_3\eta_1+\gamma_2\eta_3=\eta_2.
\label{C23}
\end{equation}
Then, since
\begin{equation}
d\theta^6=d\gamma_1\wedge dt
\label{C24}
\end{equation}
(see (\ref{C4})), Eq. (\ref{C9}) becomes
\begin{equation}
d\gamma_1\wedge dt=\eta_2\theta^4\wedge dt,
\label{C25}
\end{equation}
where Eqs. (\ref{C7}) and (\ref{C9}) have been used.\par
At this point, by elaborating the constraints (\ref{C5}) and (\ref{C6}) on 
the basis of the preceding results (\ref{C17})--(\ref{C25}) and taking 
$\theta^4=dx$, we arrive at the relations
\begin{eqnarray}
&\gamma_{2t}+\eta_{1x}-2k\gamma_2+2\gamma_1\gamma_2=0,\label{C26}\\
&\gamma_{3t}-\eta_{3x}+2k\gamma_3-2\gamma_1\gamma_3=0,\label{C27}\\
&\gamma_{1x}-\gamma_2\eta_3-\gamma_3\eta_1=0,\label{C28}\\
&\gamma_{2x}+\eta_1=0,\label{C29}\\
&\gamma_{3x}+\eta_3=0.\label{C30}
\end{eqnarray}
Integrating Eq. (\ref{C28}) with respect to $x$ and putting the function of 
integration equal to zero, with the help of Eqs. (\ref{C29}) and 
(\ref{C30}), Eqs. (\ref{C26})--(\ref{C27}) can be easily reshaped to give
\begin{eqnarray}
&\gamma_{2t}-\gamma_{2xx}-2k\gamma_2-2{\gamma_2}^2\gamma_3=0,\label{C31}\\
&\gamma_{3t}+\gamma_{3xx}+2k\gamma_3+2\gamma_2{\gamma_3}^2=0.\label{C32}
\end{eqnarray}
By identifying $\gamma_2$ and $\gamma_3$ with $-u$ and $v$, respectively, 
Eqs. (\ref{C31})--(\ref{C32}) reproduce exactly the system 
(\ref{A2a})--(\ref{A2b}). Thus, the inverse prolongation enables us to 
discover the integrable system (\ref{C26})--(\ref{C30}), closely related to 
Eqs. (\ref{A2a})--(\ref{A2b}), which can be considered as a kind of 
"potential form" of that system.

\section{The inverse prolongation via the Kac--Moody algebra}

We have shown that the incomplete Lie algebra associated with the 
prolongation structure of Eqs. (\ref{A2a})--(\ref{A2b}) allows the 
realization expressed by the Kac--Moody operators 
(\ref{B14a})--(\ref{B14c}). Here we shall exploit this realization to solve 
an inverse prolongation problem which may furnish, in theory, more 
information than the method applied in Sec. 3. To this aim, we start from the 
ansatz
\begin{equation}
\omega=dy+\left( \sum_{i=1}^{3}S_iX_i\right) dx+
\left( \sum_{i=1}^{11} \psi_{i}(S_j,\,S_{jx})X_i\right)dt,
\label{D1}
\end{equation}
where $j=1,\,2,\,3$,
\begin{eqnarray}
&X_1=\stackrel{(0)}{T_1}-i\stackrel{(0)}{T_2},\;\;\;
 X_2=\stackrel{(0)}{T_1}+i\stackrel{(0)}{T_2},\;\;\;
 X_3=\stackrel{(n)}{T_3}-\stackrel{(-n)}{T_3},\nonumber\\
&X_4=\stackrel{(n)}{T_3}+\stackrel{(-n)}{T_3},\;\;\;
 X_5=\stackrel{(0)}{T_3},\;\;\;X_6=\stackrel{(n)}{T_1},\;\;\;
 X_7=\stackrel{(-n)}{T_1},\label{D2}\\
&X_8=\stackrel{(n)}{T_2},\;\;\;X_9=\stackrel{(-n)}{T_2},\;\;\;
 X_{10}=\stackrel{(m)}{T_3},\;\;\;X_{11}=\stackrel{(-m)}{T_3},\nonumber
\end{eqnarray}
with $n,\,m\in {\cal{Z}}-\{ 0 \}$, the operators $\stackrel{(\bullet)}{T_j}$ 
being expressed by the realization (\ref{B14a})--(\ref{B14c}) of 
the Kac--Moody algebra (\ref{B13}). Then, by equating the 1--form (\ref{D1}) 
to zero, we can determine the functions $\psi_j$ and, in correspondence, 
evolution systems of the form 
$S_{jt}=f(\{S_k\},\,\{S_{kx}\},\,\{S_{kxx}\})$, where
$S_{jx}={{\partial S_j}\over{\partial x}}$, and so on.\par
In doing so, we obtain
\begin{equation}
-{\bf{y}}^{(i)}_x={1\over 2}S_1(\sigma_1+i\sigma_2){\bf{y}}^{(i)}+
{1\over 2}S_2(\sigma_1-i\sigma_2){\bf{y}}^{(i)}+
{1\over 2}S_3\sigma_3({\bf{y}}^{(n+i)}-{\bf{y}}^{(-n+i)}),
\label{D3a}
\end{equation}
\begin{eqnarray}
&-{\bf{y}}^{(i)}_t={1\over 2}\psi_1(\sigma_1+i\sigma_2){\bf{y}}^{(i)}+
 {1\over 2}\psi_2(\sigma_1-i\sigma_2){\bf{y}}^{(i)}+\nonumber\\
&+{1\over 2}\sigma_3\left[\psi_3({\bf{y}}^{(n+i)}-{\bf{y}}^{(-n+i)})+
 \psi_4({\bf{y}}^{(n+i)}+{\bf{y}}^{(-n+i)})\right]+\nonumber\\
&+{1\over 2}\psi_5\sigma_3{\bf{y}}^{(i)}+
 {1\over 2}\psi_6\sigma_1{\bf{y}}^{(n+i)}+
 {1\over 2}\psi_7\sigma_1{\bf{y}}^{(-n+i)}-
 {1\over 2}\psi_8\sigma_2{\bf{y}}^{(n+i)}-\label{D3b}\\
&-{1\over 2}\psi_9\sigma_2{\bf{y}}^{(-n+i)}+
 {1\over 2}\psi_{10}\sigma_3{\bf{y}}^{(m+i)}+
 {1\over 2}\psi_{11}\sigma_3{\bf{y}}^{(-m+i)},\nonumber
\end{eqnarray}
where $\sigma_1$, $\sigma_2$, $\sigma_3$ are the Pauli matrices acting on 
the vectors
\begin{equation}
{\bf{y}}^{(i)}=\left(\begin{array}{c}
                     y_1^{(i)} \\
                     y_2^{(i)}
          \end{array} 
                     \right).
\label{D4}
\end{equation}
The compatibility condition ${\bf{y}}^{(i)}_{xt}={\bf{y}}^{(i)}_{tx}$ for 
the system (\ref{D3a})--(\ref{D3b}) yields a set of constraints involving 
the functions $S_j$ $(j=1,\,2,\,3)$ and $\psi_k$ $(k=1,\,2,\,\ldots,11)$. 
We observe that since the index $m$ is different from $0$, $n$ and $-n$, 
two cases can occur: {\it i}) $m=2n$ (or $m=-2n$) and {\it ii}) 
$m\neq 2n,\;-2n$.\par\noindent
{\underline{\it{Case} $i$}}\par
Omitting the lengthy calculations for simplicity, for $m=2n$ we find
\begin{eqnarray}
&-S_1\psi_2+S_2\psi_1-{1\over 2}\psi_{5x}=0,\label{D5a}\\
&S_{1t}+S_1\psi_5+{1\over 2}S_3\psi_6-{1\over 2}S_3\psi_7+
 {i\over 2}S_3\psi_8-{i\over 2}S_3\psi_9-\psi_{1x}=0,\label{D5b}\\
&S_{2t}-S_2\psi_5-{1\over 2}S_3\psi_6+{1\over 2}S_3\psi_7+
 {i\over 2}S_3\psi_8-{i\over 2}S_3\psi_9-\psi_{2x}=0,\label{D5c}\\
&S_1(\psi_3+\psi_4)-S_3\psi_1-{1\over 2}\psi_{6x}-
 {i\over 2}\psi_{8x}=0,\label{D5d}\\
&-S_2(\psi_3+\psi_4)+S_3\psi_2-{1\over 2}\psi_{6x}+
 {i\over 2}\psi_{8x}=0,\label{D5e}\\
&-S_1\psi_6+iS_1\psi_8+S_2\psi_6+iS_2\psi_8+S_{3t}-
 (\psi_{3x}+\psi_{4x})=0,\label{D5f}\\
&-S_1(\psi_3-\psi_4)+S_3\psi_1-{1\over 2}\psi_{7x}-{i\over 2}
 \psi_{9x}=0,\label{D5g}\\
&S_2(\psi_3-\psi_4)-S_3\psi_2-{1\over 2}\psi_{7x}+{i\over 2}
 \psi_{9x}=0,\label{D5h}\\
&-S_1\psi_7+iS_1\psi_9+S_2\psi_7+iS_2\psi_9-S_{3t}+
 \psi_{3x}-\psi_{4x}=0,\label{D5i}\\
&S_1\psi_{10}-{1\over 2}S_3\psi_6-{i\over 2}S_3\psi_8=0,\label{D5l}\\
&-S_2\psi_{10}+{1\over 2}S_3\psi_6-{i\over 2}S_3\psi_8=0,\label{D5m}\\
&\psi_{10x}=0,\label{D5n}\\
&S_1\psi_{11}+{1\over 2}S_3\psi_7+{i\over 2}S_3\psi_9=0,\label{D5o}\\
&-S_2\psi_{11}-{1\over 2}S_3\psi_7+{i\over 2}S_3\psi_9=0,\label{D5p}\\
&\psi_{11x}=0,\label{D5q}
\end{eqnarray}
where $\psi_{jx}=\psi_{j,S_k}S_{kx}+\psi_{j,S_{kx}}S_{kxx}$. After some 
manipulations (see Appendix C), from Eqs. (\ref{D5a})--(\ref{D5q}) we are 
led to the system of nonlinear evolution equations
\begin{eqnarray}
&U_t-2aU^2V+2aU-U_x{{\psi_3}\over{S_3}}+{a\over{S_3}}
 \left({1\over {S_3}}U_x\right)_x=0,\label{D6a}\\
&V_t+2aUV^2-2aV-V_x{{\psi_3}\over{S_3}}-{a\over{S_3}}
 \left({1\over {S_3}}V_x\right)_x=0,\label{D6b}\\
&S_{3t}-\psi_{3x}=0,\label{D6c}
\end{eqnarray}
where
\begin{equation}
U={{S_1}\over {S_3}},\;\;\;V={{S_2}\over {S_3}},
\label{D7}
\end{equation}
$a$ is an arbitrary constant, and $\psi_3$ is an arbitrary function 
depending, in general, on $\{ S_j \}$ and $\{ S_{jx} \}$.\par
We remark that via the transformation
\begin{equation}
t'=t,\;\;\;x'=\alpha(x,\,t),
\label{D8}
\end{equation}
choosing the (arbitrary) function $\alpha$ in such a way that
\begin{equation}
\alpha_t=\alpha_x{{\psi_3}\over{S_3}},
\label{D9}
\end{equation}
the first derivatives $U_x$ and $V_x$ in Eqs. (\ref{D6a}) and (\ref{D6b}) 
disappear. Furthermore, assuming that
\begin{equation}
\alpha_x=S_3,\;\;\;\alpha_t=\psi_3,
\label{D10}
\end{equation}
the constraint (\ref{D6c}) is automatically satisfied and Eqs. (\ref{D6a}) and 
(\ref{D6b}) take the form
\begin{eqnarray}
&U_{t'}-2aU^2V+2aU+aU_{x'x'}=0,\label{D11a}\\
&V_{t'}+2aV^2U-2aV-aV_{x'x'}=0.\label{D11b}
\end{eqnarray}
By setting $U={{u\over {\sqrt{k}}}}$, $V={{v\over {\sqrt{k}}}}$, $a=-k$ and 
rescaling $x'$, i.e. $x'\rightarrow {1\over {\sqrt{k}}}x'$, the system 
(\ref{D11a})--(\ref{D11b}) reproduces just the original equations 
(\ref{A2a})--(\ref{A2b}).\par
To conclude, it is noteworthy that the inverse prolongation method based on 
the ansatz (\ref{D1}) and on the realization (\ref{B14a})--(\ref{B14c}) of 
the Kac--Moody algebra (\ref{B13}), is able to predict a new system of 
integrable nonlinear evolution equations, precisely Eqs. 
(\ref{D6a})--(\ref{D6c}), containing the starting equations 
(\ref{A2a})--(\ref{A2b}) as a special case.\par\noindent
{\underline{\it{Case} $ii$}}\par
For $m\neq 2n$, the compatibility condition for Eqs. (\ref{D3a})--(\ref{D3b}) 
provides the constraints
\begin{eqnarray}
&-S_{3t}+\psi_{3x}-\psi_{4x}=0,\label{D12a}\\
&S_2(\psi_3-\psi_4)-S_3\psi_2=0,\label{D12b}\\
&-S_1(\psi_3-\psi_4)+S_3\psi_1=0,\label{D12c}\\
&S_{3t}-(\psi_{3x}+\psi_{4x})=0,\label{D12d}\\
&-S_2(\psi_3+\psi_4)+S_3\psi_2=0,\label{D12e}\\
&S_{2t}-S_2\psi_5-\psi_{2x}=0,\label{D12f}\\
&S_{1t}+S_1\psi_5-\psi_{1x}=0,\label{D12g}\\
&-S_1\psi_2+S_2\psi_1-{1\over 2}\psi_{5x}=0.\label{D12h}
\end{eqnarray}
From Equations (\ref{D12b}) and (\ref{D12e}) we find
\begin{equation}
S_2\psi_4=0.\label{D13}
\end{equation}
Supposing that $S_2\neq 0$, from Eqs. (\ref{D12b})--(\ref{D12d}) we get
\begin{equation}
\psi_2={{S_2}\over{S_3}}\psi_3,\;\;\;\psi_1={{S_1}\over{S_3}}\psi_3,
\label{D14}
\end{equation}
and
\begin{equation}
S_{3t}-\psi_{3x}=0.
\label{D15}
\end{equation}
Equations (\ref{D14}) imply
\begin{equation}
S_1\psi_2-S_2\psi_1=0.
\label{D16}
\end{equation}
Taking account of (\ref{D16}), from (\ref{D12h})
\begin{equation}
\psi_{5x}=0.
\label{D17}
\end{equation}
Consequently, Eqs. (\ref{D12g}) and (\ref{D12f}) become the decoupled 
equations
\begin{eqnarray}
&U_t+\psi_5 U-{{\psi_3}\over{S_3}}U_x=0,\label{D18a}\\
&V_t-\psi_5 V-{{\psi_3}\over{S_3}}V_x=0,\label{D18b}
\end{eqnarray}
where $U={{S_1}\over {S_3}}$, $V={{S_2}\over {S_3}}$, and $S_3$ is linked 
to $\psi_3$ by the constraint (\ref{D12a}). Furthermore, the function 
$\psi_5$ depends on the variable $t$ only (see Eq. (\ref{D17})). Using the 
change of variables (\ref{D8}), Eqs. (\ref{D18a})--(\ref{D18b}) can be 
written as
\begin{equation}
U_{t'}=-\psi_5 U,\;\;\;V_{t'}=\psi_5 V,
\label{D19}
\end{equation}
which can be easily solved to give
\begin{equation}
U(x',\;t')=\beta(x')e^{-\int \psi_5(t')dt'},\;\;\;
V(x',\;t')=\beta(x')e^{\int \psi_5(t')dt'},
\label{D20}
\end{equation}
where $\beta(x')$ is a function of integration.

\section{Concluding remarks}

The analysis of the equations of the reaction--diffusion type 
(\ref{A1a})--(\ref{A1b}), in $1+1$ dimensions, carried out using the 
prolongation techniques, allows us to characterize a system, given by 
Eqs. (\ref{A2a})--(\ref{A2b}), endowed with a loop algebra structure. This 
property suggests that Eqs. (\ref{A2a})--(\ref{A2b}) constitute an 
integrable system. The direct prolongation method applied to Eqs. 
(\ref{A1a})--(\ref{A1b}) shows that a necessary condition to have an 
integrable special case, is that the diffusion coefficients $D_1$ and $D_2$ 
are of opposite sign. (We have chosen $D_1=-D_2$ for simplicity). All the 
subcases corresponding to the assumption $D_1+D_2\neq 0$ lead to 
finite--dimensional Lie algebras. This result indicates that the related 
systems are not integrable. On the other hand, apart from the integrable 
case (\ref{A2a})--(\ref{A2b}), other systems belonging to the class 
(\ref{A1a})--(\ref{A1b}) present interesting features. One of them is the 
Gierer--Meinhardt model in its simplest form \cite{K}
\begin{eqnarray}
& u_t=D_1 u_{xx}+b_1 (u^2 v-u),\label{E1a}\\
& v_t=D_2 v_{xx}+c_1 (1-u v^2),\label{E1b}
\end{eqnarray}
which admits a closed nonabelian Lie algebra connected with the similitude 
group in ${\cal R}^2$ (a subgroup of the eight--parameter projective group). 
Such an algebra (which emerges for both the choices $D_1+D_2=0$ and 
$D_1+D_2\neq 0$), and its connection with the properties of the biological 
system described by Eqs. (\ref{E1a})--(\ref{E1b}) will not be treated here. 
In this paper we have limited ourselves to carry out a systematic analysis, 
within the direct and the inverse prolongation scheme, of Eqs. 
(\ref{A2a})--(\ref{A2b}), which turn out to be a integrable particular case 
of the class of equations (\ref{A1a})--(\ref{A1b}). The prolongation Lie 
algebra related to Eqs. (\ref{A2a})--(\ref{A2b}) is incomplete and allows an 
infinite--dimensional realization of the Kac--Moody type. Moreover, the 
incomplete Lie algebra enables us to write the linear spectral problem 
associated with the system under consideration. Our results confirm that a 
close connection exists between the incomplete prolongation Lie algebra of 
the model (\ref{A2a})--(\ref{A2b}) and its integrability property.\par
On the other hand, via the inverse prolongation approach the incomplete Lie 
algebra of Eqs. (\ref{A2a})--(\ref{A2b}) is exploited to generate the field 
equations whose prolongation structure it is. The new integrable systems 
(\ref{C26})--(\ref{C27}) and (\ref{D6a})--(\ref{D6c}) are predicted, 
containing the starting equations (\ref{A2a})--(\ref{A2b}) as a special 
case.\par
To conclude, two comments are in order. First, it should be important to 
try to extend the prolongation technique to the study of higher dimension 
nonlinear field equations. In this context, although up to now some works 
have been done at the direct prolongation level \cite{M,T}, it seems that 
the inverse method has been never considered. Second, we point out that the 
correspondence between loop algebras and integrable equations is not 
unique, in the sense that the equations arising from the inverse prolongation 
depend on what we assume as independent variables.

\section*{Appendix A: the spectral problem from the Kac--Moody algebra}

Let us consider the equations $y_x=-F$, $y_t=-G$ (see (\ref{B2}), 
(\ref{B4a}) and (\ref{B4b})) with $a_1=0$. Combining together Eqs. 
(\ref{B12a})--(\ref{B12d}) and Eqs. (\ref{B14a})--(\ref{B14c}), we obtain
\begin{equation}
-y_{1,x}^{(i)}=y_2^{(i)}u+{{\sqrt{k}}\over 2}
\left[ y_1^{(n_4+i)}-y_1^{-n_4+i)}\right],
\label{ApA1}
\end{equation}
\begin{equation}
-y_{2,x}^{(i)}=y_1^{(i)}v-{{\sqrt{k}}\over 2}
\left[ y_2^{(n_4+i)}-y_2^{-n_4+i)}\right],
\label{ApA2}
\end{equation}
\begin{equation}
-y_{1,t}^{(i)}=y_2^{(i)}u_x+uvy_1^{(i)}+{\sqrt{k}}u
\left[ -y_2^{(n_4+i)}+y_2^{(-n_4+i)}\right]-{k\over 2}
\left[y_1^{(2n_4+i)}+y_1^{(-2n_4+i)}\right],
\label{ApA3}
\end{equation}
\begin{equation}
-y_{2,t}^{(i)}=y_1^{(i)}v_x-uvy_2^{(i)}+{\sqrt{k}}v
\left[-y_1^{(n_4+i)}+y_1^{(-n_4+i)}\right]+{k\over 2}
\left[y_2^{(2n_4+i)}+y_2^{(-2n_4+i)}\right].
\label{ApA4}
\end{equation}\par
Equations (\ref{ApA1})--(\ref{ApA4}) can be cast into the matrix form
\begin{eqnarray}
& -\left(
       \begin{array}{c}
               y_1^{(i)} \\
               y_2^{(i)}
       \end{array} 
                  \right)_x=
\left(
      \begin{array}{cc}
             0 & u \\
	     v & 0
      \end{array}
		\right)
\left(
       \begin{array}{c}
               y_1^{(i)} \\
               y_2^{(i)}
       \end{array} 
                  \right)+\nonumber\\
& +{{\sqrt{k}}\over 2}
\left(
      \begin{array}{cc}
             1 & 0 \\
	     0 & -1
      \end{array}
		\right)
\left[\left(
	    \begin{array}{c}
	           y_1^{(n_4+i)}\\
		   y_2^{(n_4+i)}
	    \end{array}
		      \right)-
      \left(
	    \begin{array}{c}
		   y_1^{(-n_4+i)}\\
		   y_2^{(-n_4+i)}
	    \end{array}
		      \right)
			     \right],
\label{ApA5}
\end{eqnarray}
\begin{eqnarray}
& -\left(
       \begin{array}{c}
               y_1^{(i)} \\
               y_2^{(i)}
       \end{array} 
                  \right)_t=
\left(
      \begin{array}{cc}
             uv  & u_x \\
	    -v_x & -uv
      \end{array}
		\right)
\left(
       \begin{array}{c}
               y_1^{(i)} \\
               y_2^{(i)}
       \end{array} 
                  \right)-\nonumber\\
& -{\sqrt{k}}
\left(
      \begin{array}{cc}
             0 & u \\
	     v & 0
      \end{array}
		\right)
\left[\left(
	    \begin{array}{c}
	           y_1^{(n_4+i)}\\
		   y_2^{(n_4+i)}
	    \end{array}
		      \right)-
      \left(
	    \begin{array}{c}
		   y_1^{(-n_4+i)}\\
		   y_2^{(-n_4+i)}
	    \end{array}
		      \right)
			     \right]-\label{ApA6}\\
&-{k\over 2}
\left(
      \begin{array}{cc}
	     1 & 0\\
	     0 & -1
     \end{array}
		\right)
\left[\left(
	    \begin{array}{c}
	           y_1^{(2n_4+i)}\\
		   y_2^{(2n_4+i)}
	    \end{array}
		      \right)+
    \left(
	    \begin{array}{c}
		   y_1^{(-2n_4+i)}\\
		   y_2^{(-2n_4+i)}
	    \end{array}
		      \right)
			     \right].\nonumber
\end{eqnarray}
On the other hand, by introducing the formal expansion
\begin{equation}
\psi(\epsilon)=\sum_{n=-\infty}^{+\infty}{\epsilon}^ny^{(n)}
\label{ApA7}
\end{equation}
where $\epsilon$ is a constant,
\begin{equation}
\psi(\epsilon)=\left(
                     \begin{array}{c}
        	            \psi_1(\epsilon)\\
	                    \psi_2(\epsilon)
                     \end{array}
	        \right)\;,\;\;\;
y^{(n)}=\left(
              \begin{array}{c}
	             y_1^{(n)}\\
	             y_2^{(n)}
              \end{array}  
        \right)\;,
\label{ApA8}
\end{equation}
Eqs. (\ref{ApA5})--(\ref{ApA6}) become
\begin{equation}
-\left(
       \begin{array}{c}
	      \psi_1\\
	      \psi_2
       \end{array}
 \right)_x
=\left(
       \begin{array}{cc}
	      \lambda &     u   \\
	         v    & -\lambda
       \end{array}
 \right)
\left(
      \begin{array}{c}
	     \psi_1\\
	     \psi_2
      \end{array}
\right)\;,
\label{ApA9}
\end{equation}
\begin{equation}
-\left(
       \begin{array}{c}
	      \psi_1\\
	      \psi_2
       \end{array}
 \right)_t
=\left(
       \begin{array}{cc}
	      uv-2\lambda^2-k & u_x-2\lambda u    \\
	      -v_x-2\lambda v & -(uv-2\lambda^2-k)
       \end{array}
 \right)
\left(
      \begin{array}{c}
	     \psi_1\\
	     \psi_2
      \end{array}
\right)\;,
\label{ApA10}
\end{equation}
where $\lambda={{\sqrt{k}}\over 2}(\epsilon^{-n_4}-\epsilon^{n_4})$ is the 
spectral parameter. Equations (\ref{ApA9})--(\ref{ApA10}) reproduce just the 
spectral problem (\ref{B10})--(\ref{B11}) determined starting from a 
finite--dimensional subalgebra of the incomplete prolongation algebra 
(\ref{B5a})--(\ref{B5f}).

\section*{Appendix B: the incomplete Lie algebra in terms of $A_j$}

From Eqs. (\ref{C3}) and using systematically the Jacobi identity, the 
incomplete Lie algebra (\ref{B5a})--(\ref{B5f}) takes the form
\begin{eqnarray}
 &[A_1,\,A_2]=[A_1,\,A_3]=[A_1,\,A_4]=[A_4,\,A_5]=[A_1,\,A_6]=\\ \nonumber
&=[A_1,\,A_7]=[A_1,\,A_8]=[A_3,\,A_8]=[A_2,\,A_7]=0
\label{ApB1}
\end{eqnarray}
\begin{eqnarray}
&[A_2,\,A_3]=-A_6,\;\;\;[A_2,\,A_4]=A_7,\;\;\;[A_3,\,A_4]=A_8,\\ \nonumber
&[A_2,\,A_6]=2A_2,\;\;\;[A_3,\,A_6]=-2A_3
\label{ApB2}
\end{eqnarray}
\begin{equation}
[A_1,\,A_5]=2[A_6,\,A_4]
\label{AbB3}
\end{equation}
\begin{equation}
[A_2,\,A_5]=[A_4,\,A_7]-2kA_2,
\label{ApB4}
\end{equation}
\begin{equation}
[A_3,\,A_5]=-[A_4,\,A_8]+2kA_3,
\label{ApB5}
\end{equation}
\begin{equation}
[A_7,\,A_3]=[A_8,\,A_2]+[A_4,\,A_6],
\label{ApB6}
\end{equation}
\begin{equation}
[A_6,\,A_7]=-2A_7,\;\;\;[A_6,\,A_8]=2A_8.
\label{ApB7}
\end{equation}

\section*{Appendix C: derivation of Eqs. (\ref{D6a})--(\ref{D6b})}

In order to scrutinize Eqs. (\ref{D5a})--(\ref{D5q}), let us consider first 
Eqs. (\ref{D5n}) and (\ref{D5q}), from which
\begin{equation}
\psi_{10}=a,\;\;\;\psi_{11}=b
\label{ApC1}
\end{equation}
where $a$, $b$ are supposed to be constant.\par
Then, combining together Eqs. (\ref{D5o}) and (\ref{D5q}) we have
\begin{eqnarray}
&\psi_9=-{{ib}\over{S_3}}(S_2-S_1),\label{ApC2a}\\
&\psi_7=-{b\over{S_3}}(S_1+S_2),\label{ApC2b}
\end{eqnarray}
with the help of (\ref{ApC1}).\par
Now, from Eqs. (\ref{D5l})--(\ref{D5m}) we obtain (see (\ref{ApC1}))
\begin{eqnarray}
&\psi_8={{ia}\over{S_3}}(S_2-S_1),\label{ApC3a}\\
&\psi_6={a\over{S_3}}(S_1+S_2).\label{ApC3b}
\end{eqnarray}
By virtue of (\ref{ApC2a})--(\ref{ApC3b}), Eqs. (\ref{D5g})--(\ref{D5h}) can 
be elaborated to give
\begin{equation}
(a-b)\left({{S_1S_2}\over{S_3^2}}\right)_x=0.
\label{ApC4}
\end{equation}
From Eqs. (\ref{D5a})--(\ref{D5q}), it can be shown that the only solution to 
Eq. (\ref{ApC4}) is $a=b$. Furthermore, Eqs. (\ref{ApC2a})--(\ref{ApC3b}) 
imply
\begin{equation}
\psi_6=-\psi_7,\;\;\;\psi_8=-\psi_9.
\label{ApC5}
\end{equation}
From Eqs. (\ref{D5d})--(\ref{D5e}) and (\ref{D5g})--(\ref{D5h}) we find
\begin{equation}
2(S_1+S_2)\psi_4-i(\psi_8+\psi_9)_x=0.
\label{ApC6}
\end{equation}
This constraint tells us that (see (\ref{ApC5}))
\begin{equation}
\psi_4=0,
\label{ApC7}
\end{equation}
under the assumption $S_1+S_2\neq 0$. On the other hand, we deduce
\begin{eqnarray}
&\psi_6+i\psi_8=2a{{S_1}\over {S_3}},\label{ApC8a}\\
&\psi_6-i\psi_8=2a{{S_2}\over {S_3}},\label{ApC8b}\\
&\psi_7+i\psi_9=-2a{{S_1}\over {S_3}},\label{ApC8c}\\
&\psi_7-i\psi_9=-2a{{S_2}\over {S_3}},\label{ApC8d}
\end{eqnarray}
from (\ref{ApC2a})--(\ref{ApC3b}). Thus, Eq. (\ref{D5f}) provides
\begin{equation}
S_{3t}-\psi_{3x}=0.
\label{ApC9}
\end{equation}\par
At this stage, inserting the expressions (\ref{ApC8a})--(\ref{ApC8b}) into 
Eqs. (\ref{D5d})--(\ref{D5e}) we get
\begin{equation}
\psi_3={{S_3}\over{S_1}}\psi_1+{a\over{S_1}}\left({{S_1}\over{S_3}}\right)_x=
{{S_3}\over{S_2}}\psi_2-{a\over{S_2}}\left({{S_2}\over{S_3}}\right)_x.
\label{ApC10}
\end{equation}
Equations (\ref{D5b})--(\ref{D5c}) become (see 
(\ref{ApC8a})--(\ref{ApC8d}))
\begin{eqnarray}
&S_{1t}+S_1\psi_5+2aS_1-\psi_{1x}=0,\label{ApC11a}\\
&S_{2t}-S_2\psi_5-2aS_2-\psi_{2x}=0.\label{ApC11b}
\end{eqnarray}\par
Coming back to Eq. (\ref{D5a}) and exploiting Eq. (\ref{ApC10}), we have
\begin{equation}
\psi_5=-2a{{S_1S_2}\over{S_3^2}},
\label{ApC12}
\end{equation}
up to an arbitrary function of time. Finally, substitution from 
(\ref{ApC12}) into Eqs. (\ref{ApC11a})--(\ref{ApC11b}) gives
\begin{eqnarray}
&S_{1t}-2a{{S_1^2S_2}\over{S_3^2}}+2aS_1-\psi_{1x}=0,\label{ApC13a}\\
&S_{2t}+2a{{S_1S_2^2}\over{S_3^2}}-2aS_2-\psi_{2x}=0.\label{ApC13b}
\end{eqnarray}
Putting in these equations $U={{S_1}\over{S_3}}$, $V={{S_2}\over{S_3}}$, 
and
\begin{eqnarray}
&\psi_1={{S_1}\over{S_3}}\psi_3-{a\over{S_3}}\left(
 {{S_1}\over{S_3}}\right)_x,\label{ApC14a}\\
&\psi_2={{S_2}\over{S_3}}\psi_3+{a\over{S_3}}\left(
 {{S_2}\over{S_3}}\right)_x,\label{ApC14b}
\end{eqnarray}
(see (\ref{ApC10})), we arrive at the system (\ref{D6a})--(\ref{D6b}).


\end{document}